\documentclass[twocolumn, reprint, nofootinbib]{revtex4-1}
\usepackage{amsmath, amsfonts, amssymb, array}
\usepackage{natbib}
\usepackage{mathrsfs}
\usepackage{verbatim}
\usepackage{enumerate}
\usepackage{slashed}
\usepackage[hidelinks]{hyperref}

\hypersetup{
  colorlinks   = true, 
  urlcolor     = blue, 
  linkcolor    = blue, 
  citecolor   = blue 
}

\begin{document}
\raggedbottom
\title{Relaxed hypermultiplet in four dimensional N=2 conformal supergravity}

\author{Subramanya Hegde$^{1,2}$, Bindusar Sahoo$^{1}$ and Aravindhan Srinivasan$^{1,3}$,}

\affiliation{$^{1}$Indian Institute of Science Education and Research
Thiruvananthapuram, Vithura, Kerala, 695551, India\\
$^{2}$ Harish-Chandra Research Institute, HBNI, Chhatnag Road, Jhunsi, 
Prayagraj (Allahabad), 211019 , India\\
$^{3}$ Chennai Mathematical Institute, SIPCOT IT Park, Siruseri
Kelambakkam, 603103, India} 

\begin{abstract}
Superconformal matter multiplets play a crucial role in the construction of Poincar\'e supergravity invariants. Off-shell multiplets allow for construction of general matter couplings in supergravity. In \cite{Howe:1982tm}, relaxed hypermultiplet was constructed in rigid supersymmetry which on coupling with the real scalar multiplet allowed for an off-shell formulation of the rigid hypermultiplet. In this paper, we extend the relaxed hypermultiplet to conformal supergravity. For consistency with the superconformal algebra, we find that the fields have to be allowed to transform in a non-canonical way under $SU(2)$ $R$-symmetry. We find suitable field redefinitions to obtain fields which are irreducible representations of $SU(2)$ $R$-symmetry and present the full non-linear transformation rule.
\end{abstract}

\allowdisplaybreaks
\maketitle
\section{Introduction}\label{intro}
Conformal supergravity plays an important role in constructing physical supergravity theories with higher derivative corrections. The higher degree of symmetries in conformal supergravity allows for the superconformal symmetries to be realized off-shell for certain multiplets and this plays a crucial role in the construction of higher derivative corrections to the physical supergravity theories with lesser number of symmetries than conformal supergravity. The higher derivative corrections are important for several reasons. They can be studied as supersymmetric counter-terms for removing divergences as well as anomalies of some tree level global symmetries at the loop level. They can also be used to study the higher derivative as well as quantum corrections to the entropy of black holes using techniques such as localization \cite{Dabholkar:2011ec} and entropy function \cite{Sen:2005wa,Sen:2008vm} and use the results to take the comparison with the microstate counting in string theory beyond the leading order Bekenstein-Hawking entropy.

One of the crucial ingredients of conformal supergravity is the Weyl multiplet which is a multiplet of fields that contains the graviton and its superpartner gravitino. The superconformal algebra closes on the multiplet without using any equations of motion while the structure constants of the algebra become field dependent. Such an algebra is known as a soft algebra. Further, one finds various representations of the soft algebra, which are known as matter multiplets, and consider their coupling to conformal supergravity using a set of procedures which goes by the name of superconformal multiplet calculus. Upon using some of these matter multiplets as compensators to gauge fix the additional symmetries in conformal supergravity, one gets the physical supergravity theory with symmetries belonging to the super-Poincare group. In the case of extended supergravity, one has to add auxiliary fields to the Weyl multiplet so that the off-shell degrees of freedom for bosons and fermions match. There is a standard choice of the auxiliary fields which lead to the standard Weyl multiplet (see \cite{Bergshoeff:1980is} for the $\mathcal{N}=2$ and $\mathcal{N}=4$ standard Weyl multiplet in four dimensions). There is a different choice of auxiliary fields that contains a scalar field of non vanishing mass dimension and is known as the dilaton Weyl multiplet (see \cite{Butter:2017pbp} for the $\mathcal{N}=2$ dilaton Weyl multiplet in four dimensions).

In $\mathcal{N}=2$ supergravity there are basically two matter multiplets, with spin less than $2$, which has distinct propagating degrees of freedom. The multiplet which has one propagating spin-$1$ degree of freedom with helicity $=\pm1$, two propagating spin-$0$ degrees of freedom and two propagating spin-$1/2$ degree of freedom with helicity $=\pm 1/2$ is known as the vector multiplet. Another multiplet which has four propagating spin-$0$ degrees of freedom and two propagating spin-$1/2$ degrees of freedom with helicity $=\pm 1/2$ is known as the scalar multiplet or the hypermultiplet. One can add auxiliary fields to the vector multiplet to obtain an off-shell $\mathcal{N}=2$ vector multiplet with 8+8 (bosonic+fermionic) off-shell degrees of freedom. This has been very well studied along with its coupling to $\mathcal{N}=2$ conformal supergravity and has also been used as a compensator to get $\mathcal{N}=2$ Poincare supergravity. However, to have an off-shell completion of hypermultiplet with a finite number of auxiliary fields is not very straightforward.  One of the ways  to get an off-shell completion of the $\mathcal{N}=2$ hypermultiplet is to introduce a non trivial central charge in the supersymmetry algebra \cite{Sohnius:1978fw}. Another way to get an off-shell completion of the $\mathcal{N}=2$ hypermultiplet without a central charge is to consider a different multiplet where one of the propagating spin-$0$ degree of freedom is represented by a dual 2-form gauge field. Such a multiplet is known as a linear multiplet or a tensor multiplet \cite{deWit:1979xpv, Breitenlohner:1979np}. Its off-shell completion has $8+8$ degrees of freedom. It is also very well studied along with its coupling to $\mathcal{N}=2$ conformal supergravity \cite{deWit:1982na} and has also been used as a compensator to get $\mathcal{N}=2$ Poincare supergravity. However, the dualization of a 2-form gauge field to a scalar will generally not work if we have higher number of derivatives. At the level of two derivatives, the couplings of the tensor multiplet are very restrictive in the sense that they do not couple to a spin-1 gauge field and do not have self coupling. These drawbacks of the tensor multiplet necessitates the formulation of an off-shell hypermultiplet without dualizing any spin-0 degree of freedom. Such an attempt was made in \cite{Howe:1982tm} for supersymmetric field theory, where the authors introduced two new multiplets: the $32+32$ components relaxed hypermultiplet and the $24+24$ components real scalar multiplet. They further considered a supersymmetric invariant action for the relaxed hypermultiplet and the coupling between the relaxed hypermultiplet and the real scalar multiplet. Upon using the equations of motion they showed that the combined system of the relaxed hypermultiplet and the real scalar multiplet describes the physical degrees of freedom of a hypermultiplet.

In \cite{Hegde:2017sgl}, the extension of the $24+24$ components real scalar multiplet to conformal supergravity was studied. At the linearized level, the components of this multiplet along with the $24+24$  components of an $\mathcal{N}=2$ standard Weyl multiplet coupled with the $48+48$ components current multiplet of a rigid $\mathcal{N}=2$ tensor multiplet. It was also shown that one can impose a consistent set of constraints on the real scalar multiplet to reduce its components to an $8+8$ components restricted real scalar multiplet which can be mapped to the $8+8$ components of a $\mathcal{N}=2$ tensor multiplet coupled to conformal supergravity. Further, in \cite{Hegde:2019ioy} an invariant action for the real scalar multiplet was constructed using a new density formula based on a fermionic multiplet which was used to find new higher derivative couplings of the tensor multiplet in $\mathcal{N}=2$ conformal supergravity.

In this paper, we would like to extend the $32+32$ components relaxed hypermultiplet to $\mathcal{N}=2$ conformal supergravity. This will pave the way for an off-shell treatment of the hypermultiplet in $\mathcal{N}=2$ conformal supergravity. In \cite{Butter:2015nza}, extension of relaxed hypermultiplet to conformal supergravity was studied in harmonic superspace by relating it to a well known rigid supersymmetry multiplet albeit with additional global $SU(2)$ indices. In this paper, we explictly couple the relaxed hypermultiplet to conformal supergravity. In appendix-\ref{redef}, we will comment on the relation between our results and the harmonic superspace discussion in \cite{Butter:2015nza}. 

The plan of the paper is as follows. In section-\ref{review}, we will review the relaxed hypermultiplet and real scalar multiplet in supersymmetric field theory as done in \cite{Howe:1982tm}. We will also discuss how the two multiplets were used to describe the physical degrees of freedom of a hypermultiplet. In section-\ref{ext}, we will discuss the extension of the relaxed hypermultiplet to conformal supergravity. In section-\ref{conc} we will end with some conclusions and future directions. We will also present the details of an $\mathcal{N}=2$ standard Weyl multiplet, that are relevant for the paper, in appendix-\ref{weyl}.

\section{Review of the rigid relaxed hypermultiplet}\label{review}
As we discussed in the previous section, while considering specific actions the tensor/linear multiplet provides an off-shell description of the hypermultiplet degrees of freedom using dualization, this mapping between the 2-form gauge field and the 0-form scalar field does not hold true for more general actions. 

In \cite{Howe:1982tm}, the authors tried to evade the use of 2-form gauge fields by generalizing the superspace constraints corresponding to the linear multiplet so that there is no conserved vector that can be idenitified as the field strength of a 2-form gauge field. The resulting $32+32$ multiplet was dubbed the relaxed hypermultiplet. In this section, we will elaborate on this construction and explain how when coupled with another $24+24$ multiplet, this gives the precise on-shell degrees of freedom as the hypermultiplet. As this construction was carried out in $\mathcal{N}=2$ flat space supersymmetry, components of various multiplets under consideration can be organized as irreducible representations of global $SU(2)$ symmetry. In this section, this global $SU(2)$ symmetry is simply referred to as $SU(2)$.

The superspace constraint for the linear multiplet can be written as:
\begin{align}\label{constraintlinear}
D_\alpha^{(i}\mathbf{L}^{jk)}=0,
\end{align}
where $\mathbf{L}^{ij}$ is a triplet of real superfields. Here the superspace covariant derivative $D_{\alpha i}$  is given as:
 \begin{align}\label{super_Cov_der}
D_{\alpha i}=\frac{\partial}{\partial \theta^{\alpha i}}+i\sigma^{\mu}_{\alpha\dot{\alpha}}\bar{\theta}^{\dot{\alpha}}_{i}\frac{\partial}{\partial x^{\mu}}\;,
\end{align}
where $(x^\mu,\theta^{\alpha i},\bar{\theta}_{\dot{\alpha}i})$ are the superspace coordinates written in two component Weyl spinors notation.

The above constraint in component form means that there is no field that transforms in the $\bf{4}$ representation of $SU(2)$ in the transformation of  $L^{ij}$ which is the lowest mass dimension field of the linear multiplet. That $\mathbf{L}^{ij}$ is real triplet of scalar fields means that the field $L^{ij}$ satisfies the pseuo-reality constraint,
\begin{align}
L^{ij}&=(L_{ij})^*=\varepsilon^{ik}\varepsilon^{jl}L_{kl}.
\end{align}
By an investigation of the possible $SU(2)$ representations that appear in the transformation of $L^{ij}$ and higher mass dimension fields that appear in its transformation rules, consistent with the above constraint and the supersymmetry algebra, one finds that the linear multiplet fields are $L^{ij}, G, \varphi^i$ and $H^a$. Here, $G$ is a complex scalar field and $\varphi^i$ is an $SU(2)$ doublet of Majorana fermions. Supersymmetry algebra implies that $H^a$ is a real conserved vector, which can also be seen by trying to match the bosonic and fermionic off-shell degrees of freedom. The resulting multiplet has $8+8$ off-shell degrees of freedom.

In \cite{Howe:1982tm}, a $24+24$ multiplet was presented with the superfield $\mathbf{L}^{ijkl}$ satisfying the constraint that the lowest mass dimension field $L^{ijkl}$ does not contain in its transformation any fermion that transforms in the $\bf{6}$ representation of $SU(2)$. This can be analogously written as the constraint,
\begin{align}\label{constraintcontrareal}
D_\alpha^{(i}\mathbf{L}^{jklm)}=0.
\end{align}
Analogous to $L^{ij}$, the field $L^{ijkl}$ satisfies the pseudo-reality condition,
\begin{align}
L^{ijkl}&=(L_{ijkl})^*=\varepsilon^{ip}\varepsilon^{jq}\varepsilon^{kr}\varepsilon^{ls}L_{pqrs}.
\end{align}
We will not review this multiplet here, but relevant to us is the fact that this multiplet contains a fermion $\psi^{ijk}$ that transforms in the $\bf{4}$ representation of $SU(2)$. To extend the linear multiplet to a multiplet with no conserved vectors, the authors realxed the constraint \eqref{constraintlinear} such that a fermion in $\bf{4}$ representation appears in the transformation of $L^{ij}$ and it is identified with the fermion $\psi^{ijk}$ from the multiplet that corresponds to $\mathbf{L}^{ijkl}$, while keping the constraint \eqref{constraintcontrareal} intact. Relaxation of the constraint \eqref{constraintlinear} allows for a rearrangement of the off-shell degrees of freedom and $H^a$ need not be conserved. The two multiplets coupled this way form a multiplet which contains $32+32$ off-shell degrees of freedom, with no conserved vectors. This multiplet is referred to as the relaxed hypermultiplet. Transformation rule for the components of the multiplet is given as:
\begin{widetext}
\begin{align}\label{trans-flat-relaxed-hyper}
\delta L ^ { i j } &=   \frac { 2 \sqrt{2}} { 3 } \overline { \epsilon } ^ { ( i } \lambda ^ { j ) } - \sqrt{2}\varepsilon _ { k m } \overline { \epsilon } ^ { k } \psi ^ { m i j } + \frac { 2 \sqrt{2}} { 3 } \varepsilon ^ { j k } \varepsilon ^ { i m } \overline { \epsilon } _ { ( m } \lambda _ { k ) } - \sqrt{2}\varepsilon ^ { j k } \varepsilon ^ { i m } \varepsilon ^ { l p } \overline { \epsilon } _ { l } \psi _ { p m k },\nonumber\\
\delta L ^ { i j k l } &=  \frac { 4\sqrt{2} } { 5 } \overline { \epsilon } ^ { ( i } \psi ^ { j k l )} + \frac { 4 \sqrt{2}} { 5 } \varepsilon ^ { i q } \varepsilon ^ { j r } \varepsilon ^ { k s } \varepsilon ^ { l t } \overline { \epsilon } _ { ( q } \psi _ { r s t ) },\nonumber\\
\delta \psi ^ { i j k } &=  \frac { 3 \sqrt{2}} { 8 } \epsilon ^ { ( i } M ^ { j k ) } + \frac { 3 \sqrt{2}} { 8 } \gamma ^ { \mu } \epsilon _ { m } \varepsilon ^ { m ( i } V _ { \mu } ^ { j k ) } + \frac { 5 \sqrt{2}} { 4 } \gamma ^ { \mu } \epsilon _ { m } \partial _ { \mu } L ^ { m i j k },\nonumber\\
\delta \lambda ^ { i } &=  \frac { 3\sqrt{2} } { 4 } \varepsilon _ { k j } \epsilon ^ { j } M ^ { k i } - \frac { 1 } { 4 }\sqrt{2} \epsilon ^ { i } N + \frac { 3\sqrt{2} } { 4 } \gamma ^ { \mu } \epsilon _ { j } V _ { \mu } ^ { i j } - \frac { 1 } { 4 }\sqrt{2} \varepsilon ^ { i j } \gamma ^ { \mu } \epsilon _ { j } G _ { \mu } + \frac { 3\sqrt{2} } { 2 } \gamma ^ { \mu } \epsilon _ { j } \partial _ { \mu } L ^ { i j },\nonumber\\
\delta M ^ { i j } &= \frac { 2\sqrt{2}  } { 3 } \varepsilon ^ { k ( i } \varepsilon ^ { j ) m } \overline { \xi } _ { m } \epsilon _ { k } -\frac { 8\sqrt{2} } { 3 } \partial _ { \mu } \overline { \psi } ^ { i j k } \gamma ^ { \mu } \epsilon _ { k },\nonumber\\
\delta N &= -3\sqrt{2} \varepsilon ^ { i m } \overline { \xi } _ { m } \epsilon _ { i } + 4\sqrt{2}\partial _ { \mu } \overline { \lambda } ^ { i } \gamma ^ { \mu } \epsilon _ { i },\nonumber\\
\delta V_{\mu}^{ij}&=-\frac{\sqrt{2}}{3}\varepsilon^{m(j}\bar{\epsilon}^{i)}\gamma_{\mu}\xi_{m}+\frac{4\sqrt{2}}{3}\varepsilon_{mn}\bar{\epsilon}^{n}\gamma_{\mu\nu}\partial^{\nu}\psi^{mij}-\frac{2\sqrt{2}}{3}\varepsilon_{mn}\bar{\epsilon}^{n}\partial_{\mu}\psi^{mij}\nonumber\\&\quad+\varepsilon^{im}\varepsilon^{jn}\{-\frac{\sqrt{2}}{3}\varepsilon_{k(n}\bar{\epsilon}_{m)}\gamma_{\mu}\xi^{k}+\frac{4\sqrt{2}}{3}\varepsilon^{kl}\bar{\epsilon}_{l}\gamma_{\mu\nu}\partial^{\nu}\psi_{mnk}-\frac{2\sqrt{2}}{3}\varepsilon^{kl}\bar{\epsilon}_{l} \partial_{\mu}\psi_{mnk}\},\nonumber\\
\delta G_{\mu}&=\frac{3\sqrt{2}}{2}\bar{\epsilon}^{i}\gamma_{\mu}\xi_{i}+2\sqrt{2}\varepsilon_{mi}\bar{\epsilon}^{i}\gamma_{\mu\nu}\partial^{\nu}\lambda^{m}+ \text{h.c},\nonumber\\
\delta \xi^{i}&=-\sqrt{2}\varepsilon^{ji}\varepsilon^{mk}\gamma^{\mu}\epsilon_{k}\partial_{\mu}M_{mj}+2\sqrt{2}\varepsilon^{ji}\epsilon^{k}\partial^{\mu}V_{\mu kj}+\frac{\sqrt{2}}{3}\epsilon^{i}\partial_{\mu}G^{\mu}-\sqrt{2}\varepsilon^{ji}\gamma^{\mu\nu}\epsilon^{k}\partial_{\mu}V_{\nu kj},
\end{align}
\end{widetext}
where we have given the results of \cite{Howe:1982tm} in four component formalism. The $SU(2)$ representation of the various fields of the relaxed hypermultiplet is obvious from the index structure in the above transformation rule. $M^{ij}$ and $N$ are complex triplet and singlets under $SU(2)$. $V_\mu^{ij}$ and $G_\mu$ are real triplet and singlet vector fields respectively. We have chosen the normalization of the fields in the relaxed hypermultiplet such that $[\delta_{Q}\left(\epsilon_{1}\right), \delta_{Q}\left(\epsilon_{2}\right)]X= (2 \overline{\epsilon}_{2}^{i} \gamma^{\mu} \epsilon_{1 i}+\mathrm{h.c})\partial_{\mu}X$, where $X$ is any field in the relaxed hypermultiplet. 

The above $32+32$ multiplet when coupled to a $24+24$ multiplet corresponding to a superfield contragradient to the $\bf{L^{ijkl}}$, was shown to contain the appropriate on-shell degrees of freedom as the hypermultiplet. This $24+24$ multiplet is based on a real scalar superfield $\bf{V}$ satisfying the constraint:
 \begin{align}\label{superspace-rs}
D_{\alpha\beta}\mathbf{V}=\left[D_{\alpha}^{i},\bar{D}_{\dot{\alpha}i}\right]\mathbf{V}=0\;.
\end{align}
The detailed transformation rule for the components of the multiplet in flat space was presented in \cite{Howe:1982tm} and was generalized to conformal supergravity recently in \cite{Hegde:2017sgl} following which we will refer to this multiplet as the real scalar multiplet. 

The supersymmetric invariant action with coupling between the real scalar multiplet and relaxed hyper multiplet in flat space was presented in \cite{Howe:1982tm} and is given as:
\begin{widetext}
\begin{align}\label{action}
I_1=\frac{1}{2}\int d^4x [&L_{ij}\square L^{ij}-\frac{25}{6}L_{ijkl}\square L^{ijkl}+3L_{ij}\partial^\mu V_\mu^{ij}-\frac{3}{4}M^{ij}M_{ij}+\frac{1}{18}N\bar{N}-\frac{3}{4}V^{\mu ij}V_{\mu ij}+\frac{1}{18}G^\mu G_\mu\nonumber\\
&-\frac{8}{9}\bar{\lambda}^i\slashed{\partial}\lambda_i+\frac{16}{3}\bar{\psi}^{ijk}\slashed{\partial}\psi_{ijk}-\frac{4}{3}\bar{\lambda}^i\xi_i+\text{h.c}],\nonumber\\
I_2=\int d^4x [&-\frac{1}{6}V\partial^\mu G_\mu+\frac{3}{4}K^{ij}M_{ij}+\frac{3}{4}A^{\mu ij}V_{\mu ij}+\frac{5}{8}C^{ijkl}L_{ijkl}+\bar{\psi}^i\xi_i-2\bar{\psi}^{ijk}\xi_{ijk}+\text{h.c}],
\end{align}
\end{widetext}
where $I_1$ is the kinetic action for relaxed hypermultiplet and $I_2$ is the coupling of the relaxed hypermultplet with the real scalar multiplet. $V, K^{ij}, A^{\mu ij}, C^{ijkl}, \psi^i$ and $\psi^{ijk}$ are the field components of the real scalar multiplet. $V$ is a real scalar field, $K^{ij}$ is a complex triplet of scalars. $A^{\mu ij}$ and $C^{ijkl}$ have the same pseudo reality properties as $V^{\mu ij}$ and $L^{ijkl}$.

Equation of motion for $L^{ij}$ leads to
\begin{align}
\square L^{ij}&=0.
\end{align}
Thus $L^{ij}$ correspond to three scalar on-shell degrees of freedom. 
Equations of motion for $G^\mu$ and $V^{\mu ij}$ lead to
\begin{align}
\square V &=0.
\end{align}
Thus $V$ corresponds to one scalar on-shell degree of freedom. The field $\psi_i$ is identified with $\lambda_i$ from using $\xi^i$ equations of motion. Equation of motion for $\lambda_i$ gives:
\begin{align}
\slashed{\partial}\lambda_i&=0.
\end{align}
Thus $\lambda_i$ corespond to four on-shell spin-$1/2$ degrees of freedom. The remaining equations of motion set $L^{ijkl}, C^{ijkl}, \psi^{ijk}, \xi^{ijk}$ and $\xi_i$ to zero. Thus the bosonic on-shell degrees of freedom are contained in $V, L^{ij}$ as four scalar degrees of freedom and fermionic on-shell degrees of freedom are contained in $\lambda_i$ (or $\psi_i$) as four spin-$1/2$ degrees of freedom. Thus the $56+56$ multiplet with the above action has the appropriate on-shell degrees of freedom for the $4+4$ hypermultiplet.

As mentioned earlier, the $24+24$ real scalar multiplet was generalized to $\mathcal{N}=2$ conformal supergravity in \cite{Hegde:2017sgl}. In the next section, we will generalize the $32+32$ relaxed hypermultiplet to $\mathcal{N}=2$ conformal supergravity.

\section{Extension of the $\mathcal{N}=2$ relaxed hypermultiplet to conformal supergravity}\label{ext}
In $\mathcal{N}=2$ conformal supergravity, multiplets form representations of the following soft algebra \cite{Mohaupt:2000mj}:
\begin{align}\label{soft_algebra}
    \left[\delta_{Q}\left(\epsilon_{1}\right), \delta_{Q}\left(\epsilon_{2}\right)\right]&=\delta^{(cov)}(\xi)+\delta_{M}(\varepsilon)+\delta_{K}\left(\Lambda_{K}\right)+\delta_{S}(\eta) \nonumber \\
    &\quad+\delta_{\mathrm{gauge}},\nonumber\\
     \left[\delta_{S}(\eta), \delta_{Q}(\epsilon)\right]&=\delta_{M}\left( \overline{\eta}^{i} \gamma^{a b} \epsilon_{i}+\mathrm{h.c.}\right)+\delta_{D}\left(\overline{\eta}_{i} \epsilon^{i}+\mathrm{h.c.}\right)\nonumber\\
     &\quad+\delta_{A}\left(i \overline{\eta}_{i} \epsilon^{i}+\mathrm{h.c.}\right)\nonumber
     \\ &\quad +\delta_{V}\left(-2 \overline{\eta}^{i} \epsilon_{j}-(\text { h.c.; traceless })\right),\nonumber\\
     \left[\delta_{S}\left(\eta_{1}\right), \delta_{S}\left(\eta_{2}\right)\right]&=\delta_{K}\left(\overline{\eta}_{2 i} \gamma^{a} \eta_{1}^{i}+\mathrm{h.c.}\right), 
\end{align}
where $\delta_{Q}$, $\delta_{S}$, $\delta_{M}$, $\delta_{K}$, $\delta_{D}$, $\delta_{A}$ and $\delta_{V}$ are respectively the infinitesimal transformations corresponding to Q-supersymmetry, S-supersymmetry, local Lorentz transformation, special conformal transformation, dilatation, $U(1)$ and $SU(2)$ R-symmetry. The infinitesimal transformations $\delta_{gauge}$ on the right hand side of $[\delta_{Q},\delta_{Q}]$ correspond to the additional gauge symmetries that a multiplet may possess. The infinitesimal transformation $\delta^{(cov)}(\xi)$, is the covariant general coordinate transformation defined as:
\begin{equation}
    \delta^{(cov)}(\xi)=\delta_{gct}(\xi)-\sum_{T}\delta_{T}(-\xi^{\mu}h_{\mu}(T)),
\end{equation} 
where $\delta_{gct}$ is the general coordinate transformation. In the summation, $T$ runs over all the superconformal transformations (including the additional gauge symmetries of the multiplet) except local translation, and $h_{\mu}(T)$ is the corresponding gauge field with appropriate factors as listed in table-\ref{table_superconformal} (see appendix-\ref{weyl}).
The parameters for the transformations listed on the RHS of first of the equations (\ref{soft_algebra}) are given as:
\begin{align}
    \xi^{\mu}&=2 \overline{\epsilon}_{2}^{i} \gamma^{\mu} \epsilon_{1 i}+\mathrm{h.c},\nonumber\\
    \varepsilon^{a b}&=\overline{\epsilon}_{1}^{i} \epsilon_{2}^{j} T_{i j}^{a b}+\mathrm{h.c},\nonumber\\
    \Lambda_{K}^{a}&=\overline{\epsilon}_{1}^{i} \epsilon_{2}^{j} D_{b} T_{i j}^{b a}-\frac{3}{2} \overline{\epsilon}_{2}^{i} \gamma^{a} \epsilon_{1 i} D+\mathrm{h.c},\nonumber\\
    \eta^{i}&=6 \overline{\epsilon}_{[1}^{i} \epsilon_{2]}^{j} \chi_{j}.
\end{align}
We will obtain the superconformal transformation rule for the relaxed hypermultiplet, by demanding the consistent closure of the superconformal algebra \eqref{soft_algebra} on the relaxed hypermultiplet.

The fields $L^{ij}$, $L^{ijkl}$, $V_{a}^{ij}$ and $G_{a}$ are real. Therefore, the chiral weights of these fields are $0$. Let us assign an arbitrary Weyl weight $w$ to the field $L^{ij}$. From the supersymmetry transformation rule (\ref{trans-flat-relaxed-hyper}) and from the knowledge of the weights of the supersymmetry parameters (see equation \eqref{wt_para} in appendix-\ref{weyl}), the chiral weights of the remaining fields get completely fixed, and the Weyl weights get determined in terms of $w$. The fields $L^{ij}$ and $L^{ijkl}$ are the lowest Weyl weight components in the multiplet and therefore must be invariant under S-transformations. For the S-transformation of $\psi^{ijk}$ and $\lambda^{i}$ we will consider terms allowed by the Weyl weights, chiral weights, and the global $SU(2)$ index structure as follows:
\begin{align}\label{S_transf_psi}
    \delta_{S}\lambda^{i}&= x L^{ij}\eta_{j},\nonumber\\
     \delta_{S}\psi^{ijk}&= y L^{ijkl}\eta_{l} + z \varepsilon^{l(k}L^{ij)}\eta_{l},
\end{align}
where $x$, $y$ and $z$ are arbitrary coefficients which will be determined using the $Q-S$ commutation relation. To do this we need to know how $L^{ij}$ and $L^{ijkl}$ transform under $V$-transformation. At the level of flat space supersymmetry, we had discussed in the previous section that the fields in the relaxed hypermultiplet are irreducible representations of the global $SU(2)$ symmetry. Assuming this to be true even for $SU(2)$ $R$-symmetry in conformal supergravity, we evaluate the $S-Q$ commutator on $L^{ij}$:
\begin{widetext}
\begin{align}
 [\delta_{S},\delta_{Q}]L^{ij} &=x\frac{2\sqrt{2}}{3}\overline{\epsilon}^{(i}L^{j)k}\eta_{k}-y\sqrt{2}\varepsilon_{km}\overline{\epsilon}^{k}L^{mijl}\eta_{l}-z\sqrt{2}\epsilon^{(l}L^{ij)}\eta_{l} + x\frac{2\sqrt{2}}{3}\varepsilon^{jk}\varepsilon^{im}\overline{\epsilon}_{(m}L_{k)l}\eta^{l}-y\sqrt{2}\varepsilon ^ { j k } \varepsilon ^ { i m } \varepsilon ^ { l p } \overline { \epsilon } _ { l }L_{pmkr}\eta^{r}\nonumber\\
 &\;\;\;-z\sqrt{2}\varepsilon ^ { j k } \varepsilon ^ { i m } \overline { \epsilon } _ {(r}L_{mk)}\eta^{r}\label{S_Q_LHS},
    \end{align}
 We also calculate the expected result of the commutator:
    \begin{align}
 \delta_{M}&\left( \overline{\eta}^{i} \gamma^{a b} \epsilon_{i}+\mathrm{h.c.}\right)L^{ij}+\delta_{D}\left(\overline{\eta}_{i} \epsilon^{i}+\mathrm{h.c.}\right)L^{ij}+\delta_{A}\left(i \overline{\eta}_{i} \epsilon^{i}+\mathrm{h.c.}\right)L^{ij}
    +\delta_{V}\left(-2 \overline{\eta}^{i} \epsilon_{j}-(\text { h.c.; traceless })\right)L^{ij}\nonumber\\
    &=(w-2) \bar{\eta}_{k} \epsilon^{k} L^{i j}+(w+2) \overline{\eta}^{k} \epsilon_{k} L^{ij}-4 \bar{\eta}^{(i} \epsilon_{k} L^{ j)k}+ 4 \bar{\eta}_{k} \epsilon^{(i} L^{j)k}\label{S_Q_RHS}.
\end{align}
\end{widetext}
For the superconformal algebra to close consistently equations (\ref{S_Q_LHS}) and (\ref{S_Q_RHS}) should match. But we clearly see that there is mismatch between the equations due to the presence of $L^{ijkl}$ terms in equation (\ref{S_Q_LHS}). Setting the coefficient $y=0$ might seem to resolve this mismatch; however, it leads to similar inconsistency in the $[\delta_{S},\delta_{Q}]L^{ijkl}$ equation. These mismatches that arise are rooted in the way in which relaxed hypermutliplet was constructed. The superspace constraint (\ref{constraintlinear}) was relaxed which led to the intertwining of the $24+24$ multiplet and the $8+8$ linear multiplet through the supersymmetry transformation. This suggests that for consistent coupling of the relaxed hypermultiplet to conformal supergravity, we should allow the two multiplets to get intertwined not just through Q-SUSY, but also through other superconformal transformations. Thus we come up with the following prescription to accommodate the required changes in the superconformal transformation:
\begin{enumerate}[(I)]
    \item We allow any field in the relaxed hypermultiplet to mix with any other field in the multiplet (or even the derivatives of the fields)  under V-transformation, given that such a mixing is consistent with Weyl weights, chiral weights and Lorentz structure.
    \item We also allow for covariant derivative terms consistent with the weights to appear in the S-transformation of the fields.
    \item In addition to the soft algebra relations (\ref{soft_algebra}), we make use of the following commutation relations involving V-transformations to fix the coefficients in the tranformations.
    \begin{align}
    [\delta_{V}(\lambda_{1}),\delta_{V}(\lambda_{2})]&= \delta_{V}(\lambda_{3}),\nonumber\\
    [\delta_{Q}(\epsilon),\delta_{V}(\lambda)]&=\delta_{Q}(\lambda^{j}_{k}\epsilon^{k}+h.c),\nonumber\\
    [\delta_{S}(\epsilon),\delta_{V}(\lambda)]&=\delta_{S}(-\lambda^{j}_{k}\eta_{j}+h.c),\label{V_algebra}
	\end{align}
	where $\lambda_{3\hspace{1mm}j}^{i}=\lambda_{2\hspace{1mm}k}^{i}\lambda_{1\hspace{1mm}j}^{k}-\lambda_{1\hspace{1mm}k}^{i}\lambda_{2\hspace{1mm}j}^{k}$.
\end{enumerate}

We will now illustrate our prescription. Consider the fields $L^{ij}$ and $L^{ijkl}$. They have the same Weyl weight, chiral weight and are the lowest Weyl weight components of the multiplet with no Lorentz indices. Hence we propose the following ansatz for their V-transformation:
\begin{align}
 \delta_{V}L^{ij}&= \alpha \hspace{1mm}\Lambda_{V k}^{(i}L^{j)k}+ \beta \hspace{1mm}\varepsilon_{k l} \Lambda_{V m}^{k} L^{l m i j}\label{V_transf_1}\\
 \delta_{V}L^{ijkl}&= \mu\hspace{1mm} \varepsilon^{m(l} \Lambda_{V m}^{k} L^{i j )} +\theta \hspace{1mm}\Lambda_{V m}^{(i} L^{j k l ) m}\label{V_transf_2}
\end{align}
where $\alpha$, $\beta$, $\mu$, $\theta$ are arbitrary coefficients which need to be fixed by using the soft algebra (\ref{soft_algebra}) and the commutation relations (\ref{V_algebra}). On evaluating the V-V commutator on $L^{ij}$ and $L^{ijkl}$, we find the following two possibilities for the coefficients:
\begin{enumerate}[(i)]
    \item  $\alpha=\frac{5}{2}$, $\mu\beta=\frac{3}{4}$ and $\theta=3$.
    \item $\alpha=-\frac{1}{2}$, $\mu\beta=\frac{3}{4}$ and $\theta=1$.
\end{enumerate}

On operating Q-V commutator on $L^{ij}$, we find the second possibility to be ruled out. Further by making use of the S-Q commutator on $L^{ij}$ and $L^{ijkl}$, the coefficients $\beta$, $\mu$ and $x$, $y$, $z$ get fixed. In addition the lowest Weyl weight value $w$ also gets determined.
 
\begin{align}
    \delta_{V}L^{ij}&=\frac{5}{2}\Lambda_{V k}^{(i}L^{j)k}-\frac{15}{4}\varepsilon_{kl}\Lambda_{V m}^{k}L^{lmij}\nonumber\\
     \delta_{V}L^{ijkl}&=-\frac{1}{5}\varepsilon^{m(l}\Lambda_{V m}^{k}L^{ij)}+3\Lambda_{V m}^{(i}L^{jkl)m}\nonumber\\
      \delta_{S}\psi^{ijk}&=\frac{15}{2 \sqrt{2}} L^{i j k l} \eta_{l}-\frac{1}{2 \sqrt{2}} \varepsilon^{l(k} L^{i j )} \eta_{l}\nonumber\\
     \delta_{S}\lambda^{i}&=\frac{8}{\sqrt{2}} L^{i j} \eta_{j}\nonumber\\
     w&=3
\end{align}
 Proceeding analogously, using the  aforementioned prescription and algebraic relations we determine the S- and V-transformation of all the fields in the multiplet. Having determined the S-transformation of all the fields in the multiplet, we make use of the $S-S$ commutation relation from \eqref{soft_algebra} to determine the K-transformation for fields in the multiplet. The linearized superconformal transformation rule for the relaxed hypermultiplet is found to be:
 \begin{widetext}
 \begin{align}
\delta L ^ { i j } &=   \frac { 2 \sqrt{2}} { 3 } \overline { \epsilon } ^ { ( i } \lambda ^ { j ) } - \sqrt{2}\varepsilon _ { k m } \overline { \epsilon } ^ { k } \psi ^ { m i j }  + \frac { 2 \sqrt{2}} { 3 } \varepsilon ^ { j k } \varepsilon ^ { i m } \overline { \epsilon } _ { ( m } \lambda _ { k ) } \nonumber- \sqrt{2}\varepsilon ^ { j k } \varepsilon ^ { i m } \varepsilon ^ { l p } \overline { \epsilon } _ { l } \psi _ { p m k }+ 3 \Lambda_{D}L^{ij}\nonumber\\
&\quad+ \frac{5}{2}\Lambda_{V k}^{(i}L^{j)k}-\frac{15}{4}\varepsilon_{kl}\Lambda_{V m}^{k}L^{lmij},\nonumber\\
 \delta L ^ { i j k l } & =  \frac { 4\sqrt{2} } { 5 } \overline { \epsilon } ^ { ( i } \psi ^ { j k l )} + \frac { 4 \sqrt{2}} { 5 } \varepsilon ^ { i q } \varepsilon ^ { j r } \varepsilon ^ { k s } \varepsilon ^ { l t } \overline { \epsilon } _ { ( q } \psi _ { r s t ) }+ 3 \Lambda_{D}L^{ijkl}-\frac{1}{5}\varepsilon^{m(l}\Lambda_{V m}^{k}L^{ij)}+3\Lambda_{V m}^{(i}L^{jkl)m},\nonumber\\
 \delta \psi ^ { i j k } &=  \frac { 3 \sqrt{2}} { 8 } \epsilon ^ { ( i } M ^ { j k ) }  + \frac { 3 \sqrt{2}} { 8 } \gamma ^ { a} \epsilon _ { m } \varepsilon ^ { m ( i } V _ { a } ^ { j k ) } + \frac { 5 \sqrt{2}} { 4 } \gamma ^ { a } \epsilon _ { m } \mathcal{D} _ { a} L ^ { m i j k } + \frac{7}{2} \Lambda_{D}\psi^{ijk}+\frac{i}{2}\Lambda_{A}\psi^{ijk}\nonumber\\
 &\quad+\frac{15}{2\sqrt{2}}L^{ijkl}\eta_{l}-\frac{1}{2\sqrt{2}}\varepsilon^{l(k}L^{ij)}\eta_{l}
 + 2\Lambda_{Vm}^{(i}\psi^{jk)m}-\frac{1}{6}\varepsilon^{m(i}\Lambda_{V m}^{j}\lambda^{k)}+\frac{1}{4}\Lambda_{M}^{ab}\gamma_{ab}\psi^{ijk},\nonumber\\
 \delta \lambda ^ { i } & =  \frac { 3\sqrt{2} } { 4 } \varepsilon _ { k j } \epsilon ^ { j } M ^ { k i } - \frac { 1 } { 4 }\sqrt{2} \epsilon ^ { i } N + \frac { 3\sqrt{2} } { 4 } \gamma ^ { a} \epsilon _ { j } V _ { a } ^ { i j } - \frac { 1 } { 4 }\sqrt{2} \varepsilon ^ { i j } \gamma ^ { a } \epsilon _ { j } G _ { a } + \frac { 3\sqrt{2} } { 2 } \gamma ^ { a } \epsilon _ { j } \mathcal{D}_ { a }L ^ { i j }\nonumber\\
 &\quad+\frac{7}{2} \Lambda_{D}\lambda^{i}
  +\frac{i}{2}\Lambda_{A}\lambda^{i}+\frac{8}{\sqrt{2}}L^{ij}\eta_{j}+ \frac{4}{3}\Lambda_{V j}^{i}\lambda^{j}+4\varepsilon_{jm}\Lambda_{Vk}^{m}\psi^{ijk}+\frac{1}{4}\Lambda_{M}^{ab}\gamma_{ab}\lambda^{i},\nonumber\\
     \delta M ^ { i j } &= \frac { 2\sqrt{2}  } { 3 } \varepsilon ^ { k ( i } \varepsilon ^ { j ) m } \overline { \xi } _ { m } \epsilon _ { k } -\frac { 8\sqrt{2} } { 3 } \mathcal{D} _ { a } \overline { \psi } ^ { i j k } \gamma ^ { a} \epsilon _ { k } +4\Lambda_{D}M^{ij}+i\Lambda_{A}M^{ij}-\frac{16\sqrt{2}}{3}\overline{\psi}^{ijk}\eta_{k}-\frac{4\sqrt{2}}{9}\overline{\lambda}^{(i}\varepsilon^{j)k}\eta_{k}+\Lambda_{V k}^{(i}M^{j)k}\nonumber\\
     &\quad-\frac{1}{9}\Lambda_{V k}^{(i}\varepsilon^{j)k}N,\nonumber\\
 \delta N &= -3\sqrt{2} \varepsilon ^ { i m } \overline { \xi } _ { m } \epsilon _ { i } + 4\sqrt{2}\mathcal{D} _ { a } \overline { \lambda } ^ { i } \gamma ^ { a } \epsilon _ { i }
 +4\Lambda_{D}N +i\Lambda_{A}N + \frac{12}{\sqrt{2}}\overline{\lambda}^{i}\eta_{i}-\frac{9}{2}\Lambda_{V j}^{i}\varepsilon_{ki}M^{jk},\nonumber\\
 \delta V_{a}^{ij}&=-\frac{\sqrt{2}}{3}\varepsilon^{m(j}\bar{\epsilon}^{i)}\gamma_{a}\xi_{m}+\frac{4\sqrt{2}}{3}\varepsilon_{mn}\bar{\epsilon}^{n}\gamma_{ab}\mathcal{D}^{b}\psi^{mij}-\frac{2\sqrt{2}}{3}\varepsilon_{mn}\bar{\epsilon}^{n}\mathcal{D}_{a}\psi^{mij}\nonumber\\
&\quad+\varepsilon^{im}\varepsilon^{jn}\{-\frac{\sqrt{2}}{3}\varepsilon_{k(n}\bar{\epsilon}_{m)}\gamma_{a}\xi^{k}+\frac{4\sqrt{2}}{3}\varepsilon^{kl}\bar{\epsilon}_{l}\gamma_{ab}\mathcal{D}^{b}\psi_{mnk}-\frac{2\sqrt{2}}{3}\varepsilon^{kl}\bar{\epsilon}_{l} \mathcal{D}_{a}\psi_{mnk}\}+ \frac{2\sqrt{2}}{9}(\overline{\lambda}^{(i}\gamma_{a}\eta^{j)}+\varepsilon^{ik}\varepsilon^{jm}\overline{\lambda}_{(k}\gamma_{a}\eta_{m)})\nonumber\\
&\quad-\frac{11\sqrt{2}}{3}(\overline{\psi}^{ijk}\gamma_{a}\eta^{l}\varepsilon_{kl}+\varepsilon^{iq}\varepsilon^{jr}\varepsilon^{kl}\overline{\psi}_{qrk}\gamma_{a}\eta_{l})+4\Lambda_{D}V_{a}^{ij}+ \Lambda_{V k}^{(i}V_{a}^{j)k}+\frac{1}{9}\Lambda_{V k}^{(i}\varepsilon^{j)k}G_{a}-\frac{1}{3}\Lambda_{V k}^{(i}\mathcal{D}_{a}L^{j)k}\nonumber\\
&\quad+\frac{5}{6}\varepsilon_{kl}\Lambda_{V m}^{k}\mathcal{D}_{a}L^{lmij}-\Lambda_{M a}^{b}V_{b}^{ij}+\frac{2}{3}\Lambda_{K a}L^{ij},\nonumber\\
\delta G_{a}&=\frac{3\sqrt{2}}{2}\bar{\epsilon}^{i}\gamma_{a}\xi_{i}+2\sqrt{2}\varepsilon_{mi}\bar{\epsilon}^{i}\gamma_{ab}\mathcal{D}^{b}\lambda^{m} -4\sqrt{2}\overline{\lambda}^{i}\gamma_{a}\eta^{j}\varepsilon_{ij}+ \text{h.c}+ 4\Lambda_{D}G_{a}-\frac{9}{2}\varepsilon_{kl}\Lambda_{V m}^{k}V_{a}^{lm}-\varepsilon_{kl}\Lambda_{V m}^{k}\mathcal{D}_{a}L^{lm}-\Lambda_{M a}^{b}G_{b},\nonumber\\
\delta \xi_{i}&=-\sqrt{2}\varepsilon_{ji}\varepsilon_{mk}\gamma^{a}\epsilon^{k}\mathcal{D}_{a}M^{mj}+2\sqrt{2}\varepsilon_{ji}\epsilon_{k}\mathcal{D}^{a}V_{a }^{kj}+\frac{\sqrt{2}}{3}\epsilon_{i}\mathcal{D}_{a}G^{a}
    -\sqrt{2}\varepsilon_{ji}\gamma^{ab}\epsilon_{k}\mathcal{D}_{a}V_{b}^{kj} +\frac{5}{\sqrt{2}}\gamma^{a}V_{a}^{kj}\eta_{j}\varepsilon_{ki}\nonumber\\
    &\quad+\frac{1}{3\sqrt{2}}\gamma^{a}G_{a}\eta_{i}-\frac{7}{\sqrt{2}}M^{jk}\eta^{m}\varepsilon_{km}\varepsilon_{ji}-\frac{1}{3\sqrt{2}}N\eta^{j}\varepsilon_{ji} -\frac{\sqrt{2}}{3}\gamma^{a}\mathcal{D}_{a}L^{kj}\eta_{j}\varepsilon_{ki}+\frac{9}{2}\Lambda_{D}\xi_{i}+\frac{i}{2}\Lambda_{A}\xi_{i}+\frac{1}{4}\Lambda_{M}^{ab}\gamma_{ab}\xi_{i}\nonumber\\
    &\quad-\frac{4}{9}\Lambda_{V m}^{k}\gamma^{a}\mathcal{D}_{a}\lambda^{m}\varepsilon_{ki}+\frac{4}{3}\Lambda_{V m}^{k}\gamma^{a}\mathcal{D}_{a}\psi^{mln}\varepsilon_{kl}\varepsilon_{ni}+\frac{2}{3}\Lambda_{K a}\gamma^{a}\lambda^{m}\varepsilon_{mi},
\end{align}
 \end{widetext}
where the derivative $\mathcal{D}_{a}$ that appears in the above transformations is the superconformal covariant derivative (defined by the equation (\ref{covariant_der}) in appendix-\ref{weyl}). The transformation parameters $\Lambda_{D}$, $\Lambda_{M}$, $\Lambda_{Ka}$, $\Lambda_{A}$ and $\Lambda_{V \hspace{1mm}j}^{\hspace{1mm}i}$ correspond to dilatation, local Lorentz transformation, special conformal transformation, $U(1)$ and $SU(2)$ $R$-symmetry respectively. Notice that the $S$-transformation of $\xi_{i}$ and the $V$-transformations of $V_{a}^{ij}$, $G_{a}$, and $\xi_{i}$ involves covariant derivative terms.

As was shown above, the fields in the relaxed hypermultiplet had to be allowed to transform in a non-canonical way under $SU(2)$, when we extend the multiplet to conformal supergravity. Moreover, we note in the above transformation rules that there are fields in the relaxed hypermultiplet which are not $K$-invariant as contrary to other known matter multiplets in the literature. But, we will show that by appropriate redefinition of the relaxed hypermultiplet fields we can switch to a new $SU(2)$ basis wherein all the fields transform in a way that $SU(2)$ irreps do. In addition, the fields will also be $K$-invariant in the new basis. Consider the V- transformation of $\xi_{i}$. We note that it transforms into a linear combination of $\gamma^{a} \mathcal{D}_{a} \lambda^{i}$ and $\gamma^{a} \mathcal{D}_{a} \psi^{ijk }$, and the diagonal elements (i.e terms containing $\xi_{i}$) are absent in the transformation. Thus by taking appropriate linear combinations (which we list out in appendix-\ref{redef}) of the components of the fields $\xi_{i}$, $\gamma^{a} \mathcal{D}_{a} \lambda^{i}$ and $\gamma^{a} \mathcal{D}_{a} \psi^{ijk }$ we define two new fields $\Omega$ and $\xi$  which are K-invariant and transform as $SU(2)$ scalars. Along the same lines we carry out the redefinition of the other fields as well.
\begin{table*}
			\begin{tabular}{|l | l| l | l |l |} 
				\hline
				Field \hspace{1mm}& $SU(2)$ Irreps \hspace{2mm}  &  Properties\hspace{5mm} & Weyl & Chiral \\ 
				&  &  & weight $(w)$ \hspace{2mm}& weight $(c)$\hspace{2mm}\\
				&  &  &  & \\ \hline
				&  &  &  &\\
				$A^{ijk}$ & $4$ & complex & $3$ & $0$ \\ 
				\hline
				&  &  &  &\\
				$\psi^{ij}$ & $3$ & $\gamma_{5}\psi^{ij}=\psi^{ij}$ & $7/2$ &  $1/2 $\\
				\hline
				&  &  &  &\\
				
				$\lambda^{ij}$ & $3$ & $\gamma_{5}\lambda^{ij}= \lambda^{ij}$ & $7/2$ &  $1/2 $\\
				\hline
				&  &  &  &\\
				
				$M^{i}$ & $2$ & complex & $4$ &  $1$\\
				\hline
				&  &  &  &\\
				
				$N^{i}$ & $2$ & complex & $4$ &  $1$\\
				\hline
				&  &  &  &\\
				
				$V_{a}^{i}$ & $2$ & complex, Lorentz vector & $4$ &  $0$\\
				\hline
				&  &  &  &\\
				$\xi$ & $1$ & $\gamma_{5}\xi= -\xi$ & $9/2$ &  $1/2$\\
				\hline
				&  &  &  &\\
				$\Omega$ & $1$ & $\gamma_{5}\Omega= -\Omega$ & $9/2$ &  $1/2$\\
				\hline
			\end{tabular}
			\caption{\label{table1}Field content of the relaxed hypermultiplet in the new basis}
\end{table*}

Table-\ref{table1} summarizes the field content and properties of the relaxed hypermultiplet in the new basis. The linearized transformation of the relaxed hypermultiplet in terms of the redefined fields is given as follows\footnote{See appendix-\ref{redef} for the details of the field redefiniton.}: 
\begin{widetext}
\begin{align}
\delta A^{ijk}&=-\overline{\epsilon}^{(i}\lambda^{jk)}+\varepsilon^{il}\varepsilon^{jm}\varepsilon^{kn}\overline{\epsilon}_{(l}\psi_{mn)}+3\Lambda_{D}A^{ijk}+3\Lambda_{V m}^{(i}A^{jk)m},\nonumber\\
\delta\psi^{ij}&=\epsilon^{(i}M^{j)}+\epsilon^{(i}N^{j)}+\gamma^{a}\epsilon_{m}\varepsilon^{m(i}V_{a}^{j)}-2\gamma^{a}\epsilon_{m}\mathcal{D}_{a}\overline{A}^{mij}
+ \frac{7}{2}\Lambda_{D}\psi^{ij}+\frac{i}{2}\Lambda_{A}\psi^{ij}-6\overline{A}^{ijk}\eta_{k}+2\Lambda_{V k}^{(i}\psi^{j)k}+\frac{1}{4}\Lambda_{M}^{ab}\gamma_{ab}\psi^{ij},\nonumber\\
\delta\lambda^{ij}&=\epsilon^{(i}M^{j)}-\epsilon^{(i}N^{j)}+\gamma^{a}\epsilon_{m}\varepsilon^{m(i}\varepsilon^{|k|j)}V_{a k}-2\gamma^{a}\epsilon_{m}\mathcal{D}_{a}A^{mij}
+ \frac{7}{2}\Lambda_{D}\lambda^{ij}+\frac{i}{2}\Lambda_{A}\lambda^{ij}-6A^{ijk}\eta_{k}+2\Lambda_{V k}^{(i}\lambda^{j)k}\nonumber\\
&\quad+\frac{1}{4}\Lambda_{M}^{ab}\gamma_{ab}\lambda^{ij}\nonumber\\
\delta M^{i}&= \varepsilon^{ik}\overline{\epsilon}_{k}\Omega+\overline{\epsilon}_{k}\gamma^{a}\mathcal{D}_{a}\psi^{ik}+\overline{\epsilon}_{k}\gamma^{a}\mathcal{D}_{a}\lambda^{ik}-2\overline{\eta}_{j}\psi^{ij}-2\overline{\eta}_{j}\lambda^{ij}+ 4\Lambda_{D}M^{i}+i\Lambda_{A}M^{i}+\Lambda_{V k}^{i}M^{k},\nonumber\\
\delta N^{i}&= -\varepsilon^{ik}\overline{\epsilon}_{k}\xi+\overline{\epsilon}_{k}\gamma^{a}\mathcal{D}_{a}\psi^{ik}-\overline{\epsilon}_{k}\gamma^{a}\mathcal{D}_{a}\lambda^{ik} -2\overline{\eta}_{j}\psi^{ij}+2\overline{\eta}_{j}\lambda^{ij}+ 4\Lambda_{D} N^{i}+i\Lambda_{A} N^{i}+\Lambda_{V k}^{i} N^{k},\nonumber\\
\delta V_{a}^{i}&= \frac{1}{2}\overline{\epsilon}^{i}\gamma_{a}\xi-\frac{1}{2}\overline{\epsilon}^{i}\gamma_{a}\Omega+\varepsilon_{mn}\overline{\epsilon}^{n}\gamma_{ab}\mathcal{D}^{b}\psi^{mi}-\frac{1}{3}\varepsilon_{mn}\overline{\epsilon}^{n}\mathcal{D}_{a}\psi^{mi}-\frac{1}{2}\varepsilon^{ik}\overline{\epsilon}_{k}\gamma_{a}\xi_{ L}-\frac{1}{2}\varepsilon^{ik}\overline{\epsilon}_{k}\gamma_{a}\Omega_{ L}+\varepsilon^{ik}\varepsilon^{mn}\overline{\epsilon}_{n}\gamma_{ab}\mathcal{D}^{b}\lambda_{mk}
\nonumber\\
&\quad-\frac{1}{3}\varepsilon^{ik}\varepsilon^{mn}\overline{\epsilon}_{n}\mathcal{D}_{a}\lambda_{mk}+\frac{8}{3}\overline{\eta}^{n}\gamma_{a}\psi^{im}\varepsilon_{mn}+\frac{8}{3}\overline{\eta}_{n}\gamma_{a}\lambda_{jm}\varepsilon^{mn}\varepsilon^{ij}+4\Lambda_{D}V_{a}^{i}+\Lambda_{V k}^{i}V_{a}^{k}-\Lambda_{M a}^{b}V_{b}^{i},\nonumber\\
\delta \xi&=\gamma^{a}\epsilon^{i}\mathcal{D}_{a}N^{j}\varepsilon_{ij}+\frac{3}{2}\epsilon_{k}\mathcal{D}^{a} V_{a}^{k}-\frac{3}{2}\epsilon_{k}\mathcal{D}_{a} V^{a}_{m}\varepsilon^{mk}-\frac{1}{2}\gamma^{ab}\epsilon_{k}\mathcal{D}_{a}V_{b}^{k}+\frac{1}{2}\gamma_{ab}\epsilon_{k}\mathcal{D}^{a}V^{b}_{l}\varepsilon^{lk}-4 N^{i}\eta^{j}\varepsilon_{ij}+\frac{3}{2}\gamma^{a}V_{a}^{i}\eta_{i}\nonumber\\
&\quad-\frac{3}{2}\gamma_{a}V^{a}_{k}\eta_{i}\varepsilon^{ki}+\frac{9}{2}\Lambda_{D}\xi+\frac{i}{2}\Lambda_{A}\xi+\frac{1}{4}\Lambda_{M}^{ab}\gamma_{ab}\xi,\nonumber\\
\delta \Omega&=-\gamma^{a}\epsilon^{i}\mathcal{D}_{a}M^{j}\varepsilon_{ij}-\frac{3}{2}\epsilon_{k}\mathcal{D}^{a}V_{a}^{k}-\frac{3}{2}\epsilon_{k}\mathcal{D}_{a} V^{a}_{l}\varepsilon^{lk}+\frac{1}{2}\gamma^{ab}\epsilon_{k}\mathcal{D}_{a}V_{b}^{k}+\frac{1}{2}\gamma_{ab}\epsilon_{k}\mathcal{D}^{a}V^{b}_{l}\varepsilon^{lk}+4 M^{i}\eta^{j}\varepsilon_{ij}-\frac{3}{2}\gamma^{a}V_{a}^{i}\eta_{i}\nonumber\\
&\quad-\frac{3}{2}\gamma_{a}V^{a}_{l}\eta_{i}\varepsilon^{li}+\frac{9}{2}\Lambda_{D}\Omega+\frac{i}{2}\Lambda_{A}\Omega+\frac{1}{4}\Lambda_{M}^{ab}\gamma_{ab}\Omega,
\end{align}
\end{widetext}
where $\overline{A}^{ijk}=\varepsilon^{il}\varepsilon^{jm}\varepsilon^{kn} A_{lmn}$. The subscript $L$ in $\xi_{L}$ and $\Omega_{L}$ denotes the left chirality, and the fields $\xi$ and $\Omega$ have been defined with right chirality.

Finally we obtain the complete non-linear superconformal transformation by introducing non-linear covariant terms into the  Q- and S-transformation rules. The non-linear terms are constructed by taking simple products of standard Weyl multiplet fields, curvatures and relaxed hypermultiplet fields. On careful analysis, we find that there is no such non-linear term consistent with Weyl weight, chiral weight and the representations, which can be added to the S- transformation. Hence, the S-transformation rules remain unchanged. However, for the Q-transformation there are non-linear terms consistent with weights and representations. We add all such possible terms and fix the coefficients using the algebra (\ref{soft_algebra}). We give the final result below:
\begin{widetext}
\begin{align}
\delta A^{ijk}&=-\overline{\epsilon}^{(i}\lambda^{jk)}+\varepsilon^{il}\varepsilon^{jm}\varepsilon^{kn}\overline{\epsilon}_{(l}\psi_{mn)}+3\Lambda_{D}A^{ijk}+3\Lambda_{V m}^{(i}A^{jk)m},\nonumber\\
\delta\psi^{ij}&=\epsilon^{(i}M^{j)}+\epsilon^{(i}N^{j)}+\gamma^{a}\epsilon_{m}\varepsilon^{m(i}V_{a}^{j)}-2\gamma^{a}\epsilon_{m}\mathcal{D}_{a}\overline{A}^{mij}+ \frac{7}{2}\Lambda_{D}\psi^{ij}+\frac{i}{2}\Lambda_{A}\psi^{ij}-6\overline{A}^{ijk}\eta_{k}+2\Lambda_{V k}^{(i}\psi^{j)k}\nonumber\\
&\quad+\frac{1}{4}\Lambda_{M}^{ab}\gamma_{ab}\psi^{ij},\nonumber\\
\delta\lambda^{ij}=&\epsilon^{(i}M^{j)}-\epsilon^{(i}N^{j)}+\gamma^{a}\epsilon_{m}\varepsilon^{m(i}\varepsilon^{|k|j)}V_{a k}-2\gamma^{a}\epsilon_{m}\mathcal{D}_{a}A^{mij}+ \frac{7}{2}\Lambda_{D}\lambda^{ij}+\frac{i}{2}\Lambda_{A}\lambda^{ij}-6A^{ijk}\eta_{k}+2\Lambda_{V k}^{(i}\lambda^{j)k}\nonumber\\
&\quad+\frac{1}{4}\Lambda_{M}^{ab}\gamma_{ab}\lambda^{ij},\nonumber\\
\delta M^{i}&= \varepsilon^{ik}\overline{\epsilon}_{k}\Omega+\overline{\epsilon}_{k}\gamma^{a}\mathcal{D}_{a}\psi^{ik}+\overline{\epsilon}_{k}\gamma^{a}\mathcal{D}_{a}\lambda^{ik}-2\overline{\eta}_{j}\psi^{ij}-2\overline{\eta}_{j}\lambda^{ij}+ 4\Lambda_{D}M^{i}+i\Lambda_{A}M^{i}+\Lambda_{V k}^{i}M^{k}\nonumber\\
&\quad +\frac{1}{16}\overline{\lambda}_{lj}\gamma^{ab}\epsilon_{k}\varepsilon^{kj}\varepsilon^{il}T_{ab}^{+}-\frac{1}{16}\overline{\psi}_{lj}\gamma^{ab}\epsilon_{k}\varepsilon^{kj}\varepsilon^{il}T_{ab}^{+}-9\overline{\epsilon}_{j}\chi_{k}A^{ijk}-9\overline{\epsilon}_{j}\chi_{k}\overline{A}^{ijk},\nonumber\\
\delta N^{i}&= -\varepsilon^{ik}\overline{\epsilon}_{k}\xi+\overline{\epsilon}_{k}\gamma^{a}\mathcal{D}_{a}\psi^{ik}-\overline{\epsilon}_{k}\gamma^{a}\mathcal{D}_{a}\lambda^{ik} -2\overline{\eta}_{j}\psi^{ij}+2\overline{\eta}_{j}\lambda^{ij}+ 4\Lambda_{D} N^{i}+i\Lambda_{A} N^{i}+\Lambda_{V k}^{i} N^{k}\nonumber\\
&\quad +\frac{1}{16}\overline{\lambda}_{lj}\gamma^{ab}\epsilon_{k}\varepsilon^{kj}\varepsilon^{il}T_{ab}^{+}+\frac{1}{16}\overline{\psi}_{lj}\gamma^{ab}\epsilon_{k}\varepsilon^{kj}\varepsilon^{il}T_{ab}^{+}+9\overline{\epsilon}_{j}\chi_{k}A^{ijk}-9\overline{\epsilon}_{j}\chi_{k}\overline{A}^{ijk},\nonumber\\
\delta V_{a}^{i}&= \frac{1}{2}\overline{\epsilon}^{i}\gamma_{a}\xi-\frac{1}{2}\overline{\epsilon}^{i}\gamma_{a}\Omega+\varepsilon_{mn}\overline{\epsilon}^{n}\gamma_{ab}\mathcal{D}^{b}\psi^{mi}-\frac{1}{3}\varepsilon_{mn}\overline{\epsilon}^{n}\mathcal{D}_{a}\psi^{mi} -\frac{1}{2}\varepsilon^{ik}\overline{\epsilon}_{k}\gamma_{a}\xi_{ L}-\frac{1}{2}\varepsilon^{ik}\overline{\epsilon}_{k}\gamma_{a}\Omega_{ L}+\varepsilon^{ik}\varepsilon^{mn}\overline{\epsilon}_{n}\gamma_{ab}\mathcal{D}^{b}\lambda_{mk}
\nonumber\\
&\quad -\frac{1}{3}\varepsilon^{ik}\varepsilon^{mn}\overline{\epsilon}_{n}\mathcal{D}_{a}\lambda_{mk} +\frac{8}{3}\overline{\eta}^{n}\gamma_{a}\psi^{im}\varepsilon_{mn}+\frac{8}{3}\overline{\eta}_{n}\gamma_{a}\lambda_{jm}\varepsilon^{mn}\varepsilon^{ij}+4\Lambda_{D}V_{a}^{i}+\Lambda_{V k}^{i}V_{a}^{k}-\Lambda_{M a}^{b}V_{b}^{i}\nonumber\\
&\quad -\frac{1}{12}\overline{\epsilon}^{k}\gamma^{b}\lambda_{kl}T_{ab}^{+}\varepsilon^{il}-\frac{1}{12}\overline{\epsilon}_{k}\gamma^{b}\psi^{ki}T_{ab}^{-}-3\overline{\epsilon}^{l}\gamma_{a}\chi_{j}\overline{A}^{ijm}\varepsilon_{ml}+3\overline{\epsilon}_{j}\gamma_{a}\chi^{l}\overline{A}^{ijm}\varepsilon_{ml},
\nonumber\\
\delta \xi&=\gamma^{a}\epsilon^{i}\mathcal{D}_{a}N^{j}\varepsilon_{ij}+\frac{3}{2}\epsilon_{k}\mathcal{D}^{a} V_{a}^{k}-\frac{3}{2}\epsilon_{k}\mathcal{D}_{a} V^{a}_{m}\varepsilon^{mk}-\frac{1}{2}\gamma^{ab}\epsilon_{k}\mathcal{D}_{a}V_{b}^{k}+\frac{1}{2}\gamma_{ab}\epsilon_{k}\mathcal{D}^{a}V^{b}_{l}\varepsilon^{lk}-4 N^{i}\eta^{j}\varepsilon_{ij}+\frac{3}{2}\gamma^{a}V_{a}^{i}\eta_{i}\nonumber\\
&\quad-\frac{3}{2}\gamma_{a}V^{a}_{k}\eta_{i}\varepsilon^{ki}+\frac{9}{2}\Lambda_{D}\xi+\frac{i}{2}\Lambda_{A}\xi+\frac{1}{4}\Lambda_{M}^{ab}\gamma_{ab}\xi + \frac{1}{16}\varepsilon^{ij}M_{i}\gamma^{ab}\epsilon_{j}T^{+}_{ab}-3\overline{\chi}^{i}\lambda^{jk}\epsilon_{k}\varepsilon_{ij}+3\overline{\chi}^{i}\psi^{jk}\epsilon_{k}\varepsilon_{ij}\nonumber\\
&\quad+\frac{1}{2}\gamma^{ab}\epsilon_{l}\widehat{R}_{ab}(V)^{m}_{\hspace{2mm}j}A^{jkl}\varepsilon_{mk}-\frac{1}{2}\gamma^{ab}\epsilon_{l}\widehat{R}_{ab}(V)^{m}_{\hspace{2mm}j}\overline{A}^{jkl}\varepsilon_{mk},\nonumber\\
\delta \Omega&=-\gamma^{a}\epsilon^{i}\mathcal{D}_{a}M^{j}\varepsilon_{ij}-\frac{3}{2}\epsilon_{k}\mathcal{D}^{a}V_{a}^{k}-\frac{3}{2}\epsilon_{k}\mathcal{D}_{a} V^{a}_{l}\varepsilon^{lk}+\frac{1}{2}\gamma^{ab}\epsilon_{k}\mathcal{D}_{a}{V}_{b}^{k}+\frac{1}{2}\gamma_{ab}\epsilon_{k}\mathcal{D}^{a}V^{b}_{l}\varepsilon^{lk}+4 M^{i}\eta^{j}\varepsilon_{ij}-\frac{3}{2}\gamma^{a}V_{a}^{i}\eta_{i}\nonumber\\
&\quad-\frac{3}{2}\gamma_{a}V^{a}_{l}\eta_{i}\varepsilon^{li}+\frac{9}{2}\Lambda_{D}\Omega+\frac{i}{2}\Lambda_{A}\Omega+\frac{1}{4}\Lambda_{M}^{ab}\gamma_{ab}\Omega+ \frac{1}{16}\varepsilon^{ij}N_{i}\gamma^{ab}\epsilon_{j}T^{+}_{ab}-3\overline{\chi}^{i}\lambda^{jk}\epsilon_{k}\varepsilon_{ij}-3\overline{\chi}^{i}\psi^{jk}\epsilon_{k}\varepsilon_{ij}\nonumber\\
&\quad+\frac{1}{2}\gamma^{ab}\epsilon_{l}\widehat{R}_{ab}(V)^{m}_{\hspace{2mm}j}A^{jkl}\varepsilon_{mk}+\frac{1}{2}\gamma^{ab}\epsilon_{l}\widehat{R}_{ab}(V)^{m}_{\hspace{2mm}j}\overline{A}^{jkl}\varepsilon_{mk}.\label{relax_transf_complete}
\end{align}
\end{widetext}

\section{Conclusions and future directions}\label{conc}
Matter multiplets in conformal supergravity play a crucial role in construction of physical Poincar\'e supergravity theories. Off-shell multiplets, in particular, allow the construction of general matter couplings and higher derivative invariants in supergravity which are of interest in various contexts such as higher derivative corrections to black hole entropy. 

In this paper we have extended the $32+32$ relaxed hypermultiplet constructed in \cite{Howe:1982tm} for flat space to $\mathcal{N}=2$ conformal supergravity. In doing this, we find the peculiar feature that the fields which are irreducible representations of the global $SU(2)$ symmetry must be allowed to transform in a non-canonical way under the local $SU(2)$ $R$-symmetry in order to be consistent with the superconformal algebra. We perform suitable field redefinitions so that the redefined fields are irreducible representations of the $SU(2)$ $R$-symmetry as well as invariant under $K$-symmetry. In appendix-\ref{redef}, we present the precise field redefinitions as well as the relation of our results to the constructions in harmonic superspace \cite{Butter:2015nza} and projective superspace \cite{Kuzenko:2009zu}.

In \cite{Howe:1982tm}, in flat space, the relaxed hypermultiplet was coupled to a $24+24$ real scalar multiplet to obtain an off-shell formulation for the hypermultiplet. The real scalar multiplet was extended to conformal supergravity in \cite{Hegde:2017sgl}. Further, it was shown to be reducible to an $8+8$ restricted real scalar multiplet by imposing $16+16$ constraints. This restricted real scalar multiplet admitted a tensor multiplet embedding. In \cite{Hegde:2019ioy}, a new density formula in $\mathcal{N}=2$ conformal supergravity was presented with application to real scalar multiplet. This allowed for a new higher derivative invariant for the tensor multiplet in supergravity. 

With the relaxed hypermultiplet in $\mathcal{N}=2$ conformal supergravity presented in this paper, it would be interesting to construct the analogue of the action \eqref{action} in conformal supergravity. This would allow for an off-shell formulation of hypermultiplet in four dimensional conformal supergravity. In \cite{Kuzenko:2017zsw}, an extension of the analysis from \cite{Howe:1982tm} was generalised to six dimensional $\mathcal{N}=(1,0)$ curved superspace to construct the off-shell representation for the hypermultiplet in six dimensional supergravity. This extension is different from that of this paper in the following sense. In \cite{Kuzenko:2017zsw}, an additional real superfield $\mathbf{T}$ appeared in the superspace constraints along with the $\mathbf{L}^{ij}$ and $\mathbf{L}^{ijkl}$ superfields to define the relaxed hypermultiplet. The construction of  \cite{Kuzenko:2017zsw} in six dimensions may have a straightforward generalization in four dimensions and one can have an alternate formulation of relaxed hypermultiplet with more number of components.

It will also be interesting to investigate if there exist constraints to reduce the relaxed hypermultiplet to a restricted multiplet. Whether the multiplet can be embedded into a density formula or can be mapped to other multiplets are the questions central to the techniques of superconformal tensor calculus, which need to be addressed for the case of relaxed hypermultiplet in conformal supergravity.
  
\begin{acknowledgments}
 This work is supported by SERB grant CRG/2018/002373, Govt of India. We thank Daniel Butter for useful discussion.
\end{acknowledgments}

\appendix
\section{$\mathcal{N}=2$ standard Weyl multiplet}\label{weyl}
The $\mathcal{N}=2$ standard Weyl multiplet is a superconformal gauge multiplet which has the field content $(e_{\mu}^{a}, \psi_{\mu}^{i}, b_{\mu}, A_{\mu}, \mathcal{V}_{\mu\hspace{0.1 cm}j}^{i}, T_{a b}^{i j}, \chi^{i}, D)$, with $24+24$ off-shell degrees of freedom \cite{deWit:1979dzm,Freedman:2012zz}. The fields $e_{\mu}^{a}$, $\psi_{\mu}^{i}$, $b_{\mu}$, $A_{\mu}$, $\mathcal{V}_{\mu\hspace{0.1 cm}j}^{i}$ are identified as the gauge fields of local translation ($P$), Q- supersymmetry ($Q$), dilatation ($D$), $U(1)_{R}$ symmetry ($A$) and $SU(2)_{R}$ symmetry ($V$) respectively. The fields $T_{a b}^{i j}$ which is an antiselfdual antisymmetric Lorentz tensor, $\chi^{i}$ an SU (2) doublet of Majorana spinors and $D$ a real scalar field, are auxiliary fields which are essential for the off-shell bosonic and fermionic degrees of freedom to match in the multiplet.
\begin{table*}[t]
			\begin{tabular}{|c | c| c | c | c|} 
				\hline
				Field \hspace{1mm}& $SU(2)$ Irreps \hspace{2mm}  & Weyl & Chiral  &  Chirality\\ 
				&  & weight $(w)$  & weight $(c)$ \hspace{2mm}& for \\
				&  &  &  & fermions \\ \hline
				&  &  &  &\\
				$e_{\mu}^{a}$ & $1$ & $-1$ & $0$ & \\ 
				\hline
				&  &  &  &\\
				$\psi_{\mu}^{i}$ & $2$ & $-1/2$ & $-1/2$ &  $+ $\\
				\hline
				&  &  &  &\\
				
				$b_{\mu}$ & $1$ & $0$ & $0$ &  \\
				\hline
				&  &  &  &\\
				
				$A_{\mu}$ & $1$ & $0$ & $0$ & \\
				\hline
				&  &  &  &\\
				
				$\mathcal{V}_{\mu\hspace{1mm}j}^{i}$ & $3$ & $0$ & $0$ &  \\
				\hline
				&  &  &  &\\
				
				$T^{-}_{ab}$ & $1$ & $1$ & $-1$ &  \\
				\hline
				&  &  &  &\\
				$\chi^{i}$ & $2$ & $3/2$ & $-1/2$ &  $+$\\
				\hline
				&  &  &  &\\
				$D$ & $1$ & $2$ & $0$ &  \\
				\hline
			\end{tabular}
			\caption{\label{tableWeyl} Field content for the $\mathcal{N}=2$ standard Weyl multiplet}
\end{table*}
	
Table-\ref{tableWeyl} summarizes the properties of the standard Weyl multiplet. We follow the chiral convention wherein spinors with upper $SU(2)_R$ index have positive (left) chirality and those with lower  $SU(2)_R$ index have negative (right) chirality. The chiral convention allows us to raise and lower the $SU(2)$ indices by complex conjugation. Thus for fermions complex conjugation switches the chirality. Chiral weight $(c)$ and Weyl weight ($w$) of a field $X$ are real numbers defined by the transformations:
\begin{align}
    \delta_{A}X&= ic\Lambda_{A}X,\nonumber\\
    \delta_{D}X&= w\Lambda_{D}X.\label{weight}
\end{align}
From equations (\ref{weight}) we can deduce that the chiral weight of a field and that of its complex conjugate differ by a factor of $-1$, while the Weyl weights are unaffected by complex conjugation.

The field $T^{-}_{ab}$ listed in the table-\ref{tableWeyl} is related to the field $T^{ij}_{ab}$ by:
\begin{align}
    T_{a b}^{i j}&=\frac{1}{2} T_{a b}^{-} \varepsilon^{i j}.
\end{align}
The dual of an antisymmetric Lorentz tensor is defined as:
\begin{align}
    \tilde{G}_{ab}&=\frac{1}{2}\varepsilon_{abcd}G^{cd}.
\end{align}
One can show in Minkowski space that self dual and anti selfdual parts of an antisymmetric Lorentz tensor are related by complex conjugation. Therefore, we have the following expressions relating the antiselfdual tensor $T_{ab}^{ij}$ with its selfdual counterpart:
\begin{align}
    T_{abij}&=(T_{ab}^{ij})^{*},\nonumber\\
    &=\frac{1}{2}(T_{ab}^{-}\varepsilon^{ij})^{*}=\frac{1}{2}T^{+}_{ab}\varepsilon_{ij}.
\end{align}

In addition to the gauge fields listed in the table-\ref{tableWeyl}, conformal supergravity also consists of the dependent gauge fields $\omega_{\mu}^{ab}$ corresponding to local Lorentz transformation, $f_{\mu}^{a}$ corresponding to special conformal transformation and $\phi_{\mu}^{i}$ corresponding to S-supersymmetry. These fields are determined in terms of the independent fields in the Weyl multiplet by the following set of conventional constraints:
\begin{align}
    \widehat{R}_{\mu \nu}(P)&=0,\nonumber\\
    \gamma^{\mu}\left(\widehat{R}_{\mu \nu}(Q)^{i}+\frac{1}{2}\gamma_{\mu \nu} \chi^{i}\right)&=0, \nonumber\\
   e_{b}^{\nu} \widehat{R}_{\mu \nu}(M)_{a}^{b}-i \widetilde{\widehat{R}}_{\mu a}(A)+\frac{1}{8} T_{a b i j} T_{\mu b}^{i j}-\frac{3}{2} D e_{\mu a}&=0,\label{curvature_constraints}
\end{align}
where $\widehat{R}(P)$, $\widehat{R}(Q)$, $\widehat{R}(M)$ and $\widehat{R}(A)$ are respectively the super-covariant curvatures associated with local translation, Q- supersymmetry, local Lorentz transformation and $U(1)$ R-symmetry, and $\tilde{\widehat{R}}(A)$ is the dual of $\widehat{R}(A)$. The curvature $\widehat{R}(V)$ that appears in the superconformal transformation of the relaxed hypermutliplet (\ref{relax_transf_complete}) is the super-covariant curvature associated with $SU(2)$ R-symmetry. On imposing the above constraints, the soft algebra relations (\ref{soft_algebra}) are satisfied on the fields.
\begin{table*}[t]
\begin{tabular}{|c | c| c |c| } 
 \hline
 Superconformal \hspace{1mm}&  Generator &  Normalized  & Parameter \\ 
  transformation & $T$ &  gauge field \hspace{2mm}&  \\
  &   & $h_{\mu}(T)$ &  \\ \hline
  &  &    &\\
local translation & $P^{a}$ &  $e_{\mu}^{a}$ & $\xi^{a}$ \\ 
 \hline
   &  &  &  \\
  Q-SUSY & $Q^{i}$ & $\frac{1}{2}\psi_{\mu}^{i}$ & $\epsilon^{i}$\\
 \hline
  &  &  &  \\

   dilatation& $D$ & $b_{\mu}$ & $\Lambda_{D}$   \\
 \hline
   &  &  &  \\

  $U(1)_{R}$ & $A$ & $-iA_{\mu}$ & $\Lambda_{A}$ \\
  \hline
   &  &  &  \\

   $SU(2)_{R}$ & $(V_{\Lambda})^{i}_{\hspace{1mm}j}$ & $-\frac{1}{2}\mathcal{V}_{\mu\hspace{1mm}j}^{i}$ & $\Lambda^{i}_{V\hspace{1mmj}}$   \\
  \hline
  &  &  &  \\

 S-SUSY & $S^{i}$ & $\frac{1}{2}\phi_{\mu}^{i}$ & $\eta^{i}$\\
  \hline
   &  &  &  \\
  local Lorentz & $M^{ab}$ &  $\omega_{\mu}^{ab}$ & $\varepsilon^{ab}/\Lambda_{M}^{ab}$ \\
  \hline
  &  &  &  \\
  special conformal & $K^{a}$ & $f_{\mu}^{a}$ & $\Lambda_{K}^{a}$   \\
  \hline
\end{tabular}
\caption{\label{table_superconformal}Gauge fields for superconformal transformations.}
\end{table*}
Finally we end this section by defining the supeconformal derivative and listing out the Weyl weights and chiral weights of supersymmetry parameters:
\begin{align}
    \mathcal{D}_{\mu}&= \partial_{\mu}-
    \sum_{T}\delta(h_{\mu}(T)),\nonumber\\
    \mathcal{D}_{a}&= e_{a}^{\mu}\mathcal{D}_{\mu},\label{covariant_der}
\end{align}
where the summation runs over all the superconformal generators T except local translation and $h_{\mu}(T)$ are the corresponding gauge fields as listed in table- \ref{table_superconformal}.

The chiral weight and Weyl weight of the Q- and S-SUSY parameters are:
\begin{align}
    \epsilon^{i}:&=-\frac{1}{2},\hspace{2mm}w=-\frac{1}{2},\nonumber\\
    \eta^{i}:&=-\frac{1}{2},\hspace{2mm}w=\frac{1}{2}.\label{wt_para}
\end{align}

\section{Redefiniton of the relaxed hypermultiplet fields}\label{redef}
Here we give the necessary details of redefinition of the relaxed hypermultiplet fields. Consider the V- and K- transformation of the field $\xi_{i}$:
\begin{align}
    \delta _{V}\xi_{i}&=-\frac{4}{9}\Lambda_{V m}^{k}\gamma^{a}\mathcal{D}_{a}\lambda^{m}\varepsilon_{ki}+\frac{4}{3}\Lambda_{V m}^{k}\gamma^{a}\mathcal{D}_{a}\psi^{mln}\varepsilon_{kl}\varepsilon_{ni},\nonumber\\
    \delta_{K}\xi_{i}&=\frac{2}{3}\Lambda_{K a}\gamma^{a}\lambda^{m}\varepsilon_{mi}.
\end{align}
As discussed in section-\ref{ext}, $\xi_{i}$ has a non-trivial K-transformation. Also, diagonal elements (i.e. $\xi_{i}$ terms) are absent in the V-transformation. The absence of diagonal term in the V-transformation suggests that it can be decomposed into two $SU(2)$ scalars say $\Tilde{\xi}_{1}$, $\Tilde{\xi}_{2}$. We also expect the resultant $SU(2)$ scalars to be invariant under SCT. So, we consider the following ansatz:
\begin{align}
   \Tilde{\xi}_{1/2}=&x_{1}\xi_{1} + x_{2}\xi_{2}+ x_{3}\gamma^{a}\mathcal{D}_{a}\lambda^{1}+x_{4}\gamma^{a}\mathcal{D}_{a}\lambda^{2}\nonumber\\
   &+x_{5}\gamma^{a}\mathcal{D}_{a}\psi^{111}+x_{6}\gamma^{a}\mathcal{D}_{a}\psi^{222}+x_{7}\gamma^{a}\mathcal{D}_{a}\psi^{121}\nonumber\\
   &+x_{8}\gamma^{a}\mathcal{D}_{a}\psi^{122}.\label{ansatz_1}
\end{align}
 By considering the K-transformation of the equation(\ref{ansatz_1}) and demanding $\delta_{K}\Tilde{\xi_{1}}=\delta_{K}\Tilde{\xi_{2}}=0$, we get:
\begin{align}
    \Tilde{\xi}_{1/2}=x_{1}\xi_{1} + x_{2}\xi_{2}+\frac{x_{2}}{3}\gamma^{a}\mathcal{D}_{a}\lambda^{1}-\frac{x_{1}}{3}\gamma^{a}\mathcal{D}_{a}\lambda^{2}.\label{xi_eqn}
\end{align}
By taking the V- transformations on the RHS of (\ref{xi_eqn}), one can see that the new fields are already V-invariant, i.e. the K-invariance by itself has made sure that the new fields are $SU(2)$ scalars. We have the freedom to choose the coefficients $x_{1}$ and $x_{2}$. Let us choose $x_{1}=x_{2}=1$ for $\Tilde{\xi_{1}}$, and $x_{1}=1,\text{ }  x_{2}=-1$ for $\Tilde{\xi_{2}}$:

\begin{align}
    \Tilde{\xi_{1}}=&\xi_{1} + \xi_{2}+\frac{1}{3}\gamma^{a}\mathcal{D}_{a}\lambda^{1}-\frac{1}{3}\gamma^{a}\mathcal{D}_{a}\lambda^{2},\label{eqn18}\\
    \Tilde{\xi_{2}}=&\xi_{1} - \xi_{2}-\frac{1}{3}\gamma^{a}\mathcal{D}_{a}\lambda^{1}-\frac{1}{3}\gamma^{a}\mathcal{D}_{a}\lambda^{2}.
\end{align}
So, in effect we have decomposed the two $SU(2)$ components of $\xi_{i}$ into $1\oplus1$ of $SU(2)$.

Next we look at $M^{ij}$ and $N$. These fields mix with each other under $SU(2)$ and have a total of 4 $SU(2)$ components. So, with these four components, the possible irreps of $SU(2)$ that we can form are $4$, $3\oplus1$ and $2\oplus2$. 4 of $SU(2)$ has 3 fundamental $SU(2)$ indices, 3 of $SU(2)$ has 2, 2 of $SU(2)$ has 1 and 1 of SU(2) has 0 $SU(2)$ indices. We know that Q- SUSY transformations of $M^{ij}$ and $N$ contain $\xi_{i}$ terms. Therefore, the new irreps that we form out of $M^{ij}$ and $N$ should contain $\Tilde{\xi}_{1/2}$ terms in the Q- transformations. For these transformations to be consistent with the index structure, one has to decompose $M^{ij}$ and $N$ as $2\oplus2$. Similar arguments follow for the remaining fields. Table-\ref{table2} summarizes the decomposition of all the fields.
\begin{table*}
 \begin{tabular}{|c| c| c | c  |} 
 \hline
 Old fields & No. of $SU(2)$ components &  Decomposition & New fields\\  
 \hline
   &    &   &\\
 $\xi_{i}$ & 2& $1\oplus1$ & $\Tilde{\xi}_{1}, \Tilde{\xi}_{2}$ \\ 
 \hline
 &    &   &\\
 $M^{ij},N$ & 4 & $2\oplus2$ & $\Tilde{M^{i}},\Tilde{N^{i}} $  \\
 \hline
 &    &   &\\
 $V_{a}^{ij},G_{a}$& 4 & $2\oplus2$ & $\Tilde{V}_{a}^{i},\Tilde{G}_{a}^{i} $ \\
 \hline
  &    &   &\\
 $\lambda^{i},\psi^{ijk}$& 6 & $3\oplus3$ & $\Tilde{\lambda}^{ij},\Tilde{\psi}^{ij}$\\
 \hline
 &    &   &\\
  $L^{ij},L^{ijkl}$& 8 & $4\oplus4$ & $\Tilde{A}^{ijk},\Tilde{B}^{ijk}$\\
 \hline
\end{tabular}
 \caption{Field redefinitions to obtain the components as $SU(2)_R$ irreps}
    \label{table2}
\end{table*}
We list out the remaining field redefintions below:
\begin{widetext}
\begin{align}
\left(\tilde{M}^{1}, \tilde{M}^{2}\right)&=\left(M^{11}+M^{12}+\frac{N}{9}, M^{12}+M^{22}-\frac{N}{9}\right),\nonumber\\
\left(\tilde{N}^{1}, \tilde{N}^{2}\right)&=\left(M^{11}-M^{12}-\frac{N}{9}, M^{12}-M^{22}-\frac{N}{9}\right),\nonumber\\
(\tilde{V_{a}}^{1},\tilde{V_{a}}^{2})&=(V_{a}^{11}+\frac{2}{9}\mathcal{D}_{a}L^{11},V_{a}^{12}+\frac{1}{9}G_{a}+\frac{2}{9}\mathcal{D}_{a}L^{12}),\nonumber\\
(\tilde{G_{a}}^{1},\tilde{G_{a}}^{2})&=(V_{a}^{12}-\frac{1}{9}G_{a}+\frac{2}{9}\mathcal{D}_{a}L^{12},V_{a}^{22}+\frac{2}{9}\mathcal{D}_{a}L^{22}),\nonumber\\
(\tilde{\psi}^{11},\tilde{\psi}^{12},\tilde{\psi}^{22})&=(\psi^{111},\psi^{112}+\frac{1}{12}\lambda^{1},\psi^{122}+\frac{1}{6}\lambda^{2}),\nonumber\\
(\tilde{\lambda}^{11},\tilde{\lambda}^{12},\tilde{\lambda}^{22})&=(\psi^{112}-\frac{1}{6}\lambda^{1},\psi^{122}-\frac{1}{12}\lambda^{2},\psi^{222}),\nonumber\\
(\tilde{A}^{111},\tilde{A}^{112},\tilde{A}^{122},\tilde{A}^{222})&=(L^{11}-5L^{1112},\frac{2}{3}L^{12}-5L^{1122},\frac{1}{3}L^{22}-5L^{1222},-5L^{2222}),\nonumber\\
(\tilde{B}^{111},\tilde{B}^{112},\tilde{B}^{122},\tilde{B}^{222})&=(L^{1111},\frac{1}{15}L^{11}+L^{1112},\frac{2}{15}L^{12}+L^{1122},\frac{1}{5}L^{22}+L^{1222}).
\end{align}
\end{widetext}

In the new basis, the reality constraints of $L^{ij}$, $L^{ijkl}$, $V_{a}^{ij}$ and $G_{a}$ translate into:
\begin{align}
\Tilde{B}^{ijk}=&-\frac{1}{5}\varepsilon^{il}\varepsilon^{jm}\varepsilon^{kn}(\Tilde{A}^{lmn})^{*},\nonumber\\
\Tilde{G}_{a}^{i}=&\varepsilon^{ji}(\Tilde{V}_{a}^{j})^{*}.
\end{align}
Further to tidy up things, we make the following normalization choice for the fields:
\begin{align}
    A^{ijk}&=\Tilde{A}^{ijk} ,(B^{ijk}=\Tilde{B}^{ijk}),\nonumber\\
    \lambda^{ij}&=4\sqrt{2}\Tilde{\lambda}^{ij},\nonumber\\
    \psi^{ij}&=4\sqrt{2}\Tilde{\psi}^{ij},\nonumber\\
    M^{i}&=\frac{3}{2}\Tilde{M}^{i},\nonumber\\
    N^{i}&=\frac{3}{2}\Tilde{N}^{i},\nonumber\\
    V_{a}^{i}&=3\Tilde{V}_{a}^{i},\nonumber\\
    \xi&=\sqrt{2}\Tilde{\xi}_{1},\nonumber\\
    \Omega&=\sqrt{2}\Tilde{\xi}_{2}.
\end{align}
Also we define $\overline{A}^{ijk}=\varepsilon^{il}\varepsilon^{jm}\varepsilon^{kn} A_{lmn}$, which implies $B^{ijk}=-\frac{1}{5}\overline{A}^{ijk}$.

We will now discuss the relation between our results and the discussion in \cite{Butter:2015nza}\footnote{We thank Daniel Butter for useful correspondence regarding this.}. In Table-\ref{table2}, we see that for each mass dimension we have two sets of fields which transform under the same representation of $SU(2)$. At the the lowest mass dimension level, we have $A^{ijk}$ and $B^{ijk}$ fields that transform under $\bf{4}$ representation of $SU(2)$ $R$-symmetry. In \eqref{relax_transf_complete}, we see that in the transformation rule for $A^{ijk}$ there is no fermionic field that transforms under $\bf{5}$ representation of $SU(2)$. We can write this constraint in superspace, analogous to the discussion in section-\ref{review}, as follows:
\begin{align}\label{superfield-O3}
D_\alpha^{(i}\mathbf{L}^{jkl)a}=0,
\end{align}
where $a$ runs over $1,2$ and is an additional global $SU(2)$ index carried by the superfield $\mathbf{L}^{ijka}$ whose lowest mass dimension components are $L^{ijka}=(A^{ijk},B^{ijk})$. At each mass dimension, the two fields with the same representation under $SU(2)$ $R$-symmetry are now placed inside a $\bf{2}$ representation of the additional global $SU(2)$ using the index $a$, upto field redefinitions. Clearly, the field redefinitions carried out in the present section are valid even in the case of rigid supersymmetry. Therefore at the level of rigid supersymmetry, relaxed hypermultiplet can be identified as the superfield $\mathbf{L}^{ijka}$. Such an identification was made at the rigid supersymmetry level in \cite{Butter:2015nza}. The superfield $\mathbf{L}^{ijka}$ was identified as the well known $O(3)$ superfield with an additional flavor index $a$. The $O(3)$ superfield falls under the general class of $O(n)$ superfields defined by the constraint:
\begin{align}\label{superfield-On}
D_\alpha^{(i_0}\mathbf{L}^{i_1i_2...i_n)}=0.
\end{align}
In this language, the linear multiplet defined by the constraint \eqref{constraintlinear} is referred to as the $O(2)$ multiplet. Coupling of $O(n)$ multiplets to conformal supergravity was discussed in \cite{Butter:2015nza} in harmonic superspace and in \cite{Kuzenko:2009zu} in projective superspace.

Thus in \cite{Butter:2015nza}, it was argued that extension of relaxed hypermultiplet to conformal supergravity would be the same as the $O(3)$ superfield. In this paper, we explicitly coupled the relaxed hypermultiplet to conformal supergravity and indeed, we find that the only way to extend it to conformal supergravity is through an $O(3)$ superfield with an additional flavor $SU(2)$ index.

\bibliography{references}

\providecommand{\href}[2]{#2}\begingroup\raggedright\begin{thebibliography}{10}

\bibitem{Howe:1982tm}
P.~S. Howe, K.~S. Stelle, and P.~K. Townsend, ``{The Relaxed Hypermultiplet: An
  Unconstrained N=2 Superfield Theory},''
\href{http://dx.doi.org/10.1016/0550-3213(83)90249-3}{{\em Nucl. Phys.}
  {\bfseries B214} (1983) 519--531}.

\bibitem{Dabholkar:2011ec}
A.~Dabholkar, J.~Gomes, and S.~Murthy, ``{Localization \& Exact Holography},''
  \href{http://dx.doi.org/10.1007/JHEP04(2013)062}{{\em JHEP} {\bfseries 04}
  (2013) 062},
\href{http://arxiv.org/abs/1111.1161}{{\ttfamily arXiv:1111.1161 [hep-th]}}.

\bibitem{Sen:2005wa}
A.~Sen, ``{Black hole entropy function and the attractor mechanism in higher
  derivative gravity},''
  \href{http://dx.doi.org/10.1088/1126-6708/2005/09/038}{{\em JHEP} {\bfseries
  09} (2005) 038},
\href{http://arxiv.org/abs/hep-th/0506177}{{\ttfamily arXiv:hep-th/0506177
  [hep-th]}}.

\bibitem{Sen:2008vm}
A.~Sen, ``{Quantum Entropy Function from AdS(2)/CFT(1) Correspondence},''
  \href{http://dx.doi.org/10.1142/S0217751X09045893}{{\em Int. J. Mod. Phys.}
  {\bfseries A24} (2009) 4225--4244},
\href{http://arxiv.org/abs/0809.3304}{{\ttfamily arXiv:0809.3304 [hep-th]}}.

\bibitem{Bergshoeff:1980is}
E.~Bergshoeff, M.~de~Roo, and B.~de~Wit, ``{Extended Conformal Supergravity},''
\href{http://dx.doi.org/10.1016/0550-3213(81)90465-X}{{\em Nucl. Phys.}
  {\bfseries B182} (1981) 173--204}.

\bibitem{Butter:2017pbp}
D.~Butter, S.~Hegde, I.~Lodato, and B.~Sahoo, ``{$N=2$ dilaton Weyl multiplet
  in 4D supergravity},'' \href{http://dx.doi.org/10.1007/JHEP03(2018)154}{{\em
  JHEP} {\bfseries 03} (2018) 154},
\href{http://arxiv.org/abs/1712.05365}{{\ttfamily arXiv:1712.05365 [hep-th]}}.

\bibitem{Sohnius:1978fw}
M.~F. Sohnius, ``{Supersymmetry and Central Charges},''
\href{http://dx.doi.org/10.1016/0550-3213(78)90159-1}{{\em Nucl. Phys.}
  {\bfseries B138} (1978) 109--121}.

\bibitem{deWit:1979xpv}
B.~de~Wit and J.~W. van Holten, ``{Multiplets of Linearized SO(2)
  Supergravity},''
\href{http://dx.doi.org/10.1016/0550-3213(79)90285-2}{{\em Nucl. Phys.}
  {\bfseries B155} (1979) 530--542}.

\bibitem{Breitenlohner:1979np}
P.~Breitenlohner and M.~F. Sohnius, ``{Superfields, Auxiliary Fields, and
  Tensor Calculus for $N=2$ Extended Supergravity},''
\href{http://dx.doi.org/10.1016/0550-3213(80)90045-0}{{\em Nucl. Phys.}
  {\bfseries B165} (1980) 483--510}.

\bibitem{deWit:1982na}
B.~de~Wit, R.~Philippe, and A.~Van~Proeyen, ``{The Improved Tensor Multiplet in
  $N=2$ Supergravity},''
\href{http://dx.doi.org/10.1016/0550-3213(83)90432-7}{{\em Nucl. Phys.}
  {\bfseries B219} (1983) 143--166}.

\bibitem{Hegde:2017sgl}
S.~Hegde, I.~Lodato, and B.~Sahoo, ``{24+24 real scalar multiplet in four
  dimensional N=2 conformal supergravity},''
  \href{http://dx.doi.org/10.1103/PhysRevD.97.066026}{{\em Phys. Rev.}
  {\bfseries D97} no.~6, (2018) 066026},
\href{http://arxiv.org/abs/1712.02309}{{\ttfamily arXiv:1712.02309 [hep-th]}}.

\bibitem{Hegde:2019ioy}
S.~Hegde and B.~Sahoo, ``{New higher derivative action for tensor multiplet in
  N=2 conformal supergravity in four dimensions},''
\href{http://arxiv.org/abs/1911.09585}{{\ttfamily arXiv:1911.09585 [hep-th]}}.

\bibitem{Butter:2015nza}
D.~Butter, ``{On conformal supergravity and harmonic superspace},''
  \href{http://dx.doi.org/10.1007/JHEP03(2016)107}{{\em JHEP} {\bfseries 03}
  (2016) 107},
\href{http://arxiv.org/abs/1508.07718}{{\ttfamily arXiv:1508.07718 [hep-th]}}.

\bibitem{Mohaupt:2000mj}
T.~Mohaupt, ``{Black hole entropy, special geometry and strings},''
  \href{http://dx.doi.org/10.1002/1521-3978(200102)49:1/3<3::AID-PROP3>3.0.CO;2-#}{{\em
  Fortsch. Phys.} {\bfseries 49} (2001) 3--161},
\href{http://arxiv.org/abs/hep-th/0007195}{{\ttfamily arXiv:hep-th/0007195
  [hep-th]}}.

\bibitem{Kuzenko:2009zu}
S.~M. Kuzenko, U.~Lindstrom, M.~Rocek, and G.~Tartaglino-Mazzucchelli, ``{On
  conformal supergravity and projective superspace},''
  \href{http://dx.doi.org/10.1088/1126-6708/2009/08/023}{{\em JHEP} {\bfseries
  08} (2009) 023},
\href{http://arxiv.org/abs/0905.0063}{{\ttfamily arXiv:0905.0063 [hep-th]}}.

\bibitem{Kuzenko:2017zsw}
S.~M. Kuzenko, J.~Novak, and S.~Theisen, ``Non-conformal supercurrents in six
  dimensions,'' {\em JHEP} {\bfseries 02} (2018) 030.

\bibitem{deWit:1979dzm}
B.~de~Wit, J.~W. van Holten, and A.~Van~Proeyen, ``{Transformation Rules of N=2
  Supergravity Multiplets},''
\href{http://dx.doi.org/10.1016/0550-3213(80)90125-X}{{\em Nucl. Phys.}
  {\bfseries B167} (1980) 186}.

\bibitem{Freedman:2012zz}
D.~Z. Freedman and A.~Van~Proeyen, {\em {Supergravity}}.
\newblock Cambridge Univ. Press, Cambridge, UK, 2012.
\newblock
\url{http://www.cambridge.org/mw/academic/subjects/physics/theoretical-physics-and-mathematical-physics/supergravity?format=AR}.
\newblock

\end{thebibliography}\endgroup
\bibliographystyle{utphys}

\end{document}